\documentclass[12pt,a4paper]{article}
\usepackage{overcite}
\usepackage{graphicx}
\usepackage{wasysym}

\usepackage[usenames]{color}
\usepackage[normalem]{ulem}

\newcommand{\lb}{\lambda_{\rm B}}
\newcommand{\dr}[1]{_{\rm #1}}
\newcommand{\ur}[1]{^{\rm #1}}

\definecolor{DarkBlue} {rgb}{0,0.08,0.45}
\definecolor{DarkGreen}{rgb}{0,0.45,0.08}
\definecolor{DarkRed}  {rgb}{0.45,0.08,0}

\newcommand{\noi}{\noindent}

\long\def\sblfootnote[#1]#2{\begingroup%
\def\thefootnote{\fnsymbol{footnote}}\footnote[#1]{#2}\endgroup}

\begin{document}
 % \pagenumbering{no}
 % \pagestyle{myheadings}
 % \setcounter{page}{1}

{\center
{\large \bf Phase Phenomena in Supported Lipid Films \\Under Varying Electric Potential \\} %: Monte Carlo Simulations \\and Mean-Field Calculations \\}
\vspace{1cm}
  {\bf  Andrey V. Brukhno\sblfootnote[1]{Corresponding author, E-mail: a.brukhno@leeds.ac.uk}\sblfootnote[2]{Also known as Andrei V. Broukhno}$^{1}$, 
%                                                                                                E-mail: a.brukhno@leeds.ac.uk, abrukhno@gmail.com}, 
        Anna Akinshina$^{2}$, Zachary Coldrick$^{1}$, \\ Andrew Nelson$^{1}$ and Stefan Auer$^{1}$\\
	\vspace{0.5cm}
	{\small \it $^{1}$ Centre for Molecular Nanoscience (CMNS), University of Leeds, United Kingdom \\
	$^{2}$ Unilever R\&D, Port Sunlight, Wirral, CH63 3JW, United Kingdom \\
	}

  }

}

\vskip 30pt

\abstract{
We model cyclic voltammetry experiments on supported lipid films where a non-trivial dependence of the capacitance on 
the applied voltage is observed. Previously, based on a mean-field treatment of the Flory-Huggins type, under the assumption 
of strongly screened electrostatic interactions, it has been hypothesized that peaks in the capacitance-vs-voltage profiles 
correspond to a sequence of structural or phase transitions within the interface. To examine this hypothesis, in this study 
we use both mean-field calculations and Monte Carlo simulations where the electrostatic effects due to the varying electric 
potential and the presence of salt are accounted for explicitly. 
Our main focus is on the structure of the film and the desorption--readsorption phenomena. These are found to be driven by 
a strong competition for the progressively charged-up (hydrophobic) surface between lipid hydrocarbon tails and the electrode 
counterions (cations). 
As the surface charge density is raised, the following phase phenomena within the interface are clearly observed: 
(i) a gradual displacement of the monolayer from the surface by the counterions, leading to  complete monolayer desorption 
and formation of an electric double layer by the surface, 
(ii) a transformation of the monolayer into a bilayer upon its desorption, 
(iii) in the case of zwitterionic (or strongly polar) lipid head groups, the desorption is followed by the bilayer 
readsorption to the electrode via interaction with the electric double layer and release of the excess 
counterions into the bulk solution. We argue then that the voltammetry peaks are associated with a stepwise process 
of formation of layers of alternating charge: electric double layer -- upon film desorption, triple or multi-layer 
-- upon film readsoption. 
}

\vfill
\newpage

\section{Introduction} 

Supported phospholipid monolayers, being significantly easier to manipulate in experiment than biological membranes, 
are often used as a convenient model for those. Indeed, monolayers have been a ``working horse'' in numerous studies 
of phospholipid bilayers.~\cite{Seul-etal:83,FAML-LAN:90,LAN-FAML:90,Tamm-etal:01,Konovalove-etal:02,Jensen-etal:04,We-Lee:09,Coldrick-etal:09,Becucci-etal-Guidelli:03,Becucci-Guidelli:09,Becucci-etal-Guidelli:10} 
One way of characterising layered nanoscopic films and their interactions with the surrounding electrolyte solution 
is via electrochemical experiments. A typical electrochemical setup -- in voltammetry, potentiometry, impedance spectroscopy, 
and such, is with phospholipids, or other surface active compounds, self-assembled into layered structures by adsorption 
onto a planar, preferably atomically smooth (e.g. gold-coated or mercury-droplet), electrode to which a varying electric 
potential is applied. Such experiments are based on measuring conductivity, charge capacity and ion-permeability of adsorbed 
material. The aim usually is either (i) studying the interfacial reactions and structural transformations in response to varying 
electric field,~\cite{LAN-FAML:90,Nelson:91,Bizzotto-Nelson:98,Conway-edts:99,Monne-etal-Nelson:03,Monne-etal-Nelson:07,Limon:08,Becucci-Guidelli:09,Becucci-etal-Guidelli:10} 
or (ii) using the electrochemical phenomena in molecular sensing devices, e.g. bio- and taste 
sensors.~\cite{Seul-etal:83,Decher:97,Konovalove-etal:02,Jensen-etal:04,SantosJr-etal:03} In spite of a broad application range and great 
investigation effort, the structural and phase behaviour as well as the details of molecular interactions with/within supported 
nano-films still lack a comprehensive understanding and, hence, the reliable ways of monitoring and controlling these phenomena. 
This is one of the areas where numerical modelling can provide invaluable insights into the molecular %structure and transformation 
mechanisms~\cite{FAML-LAN:90,Leermakers:88,Cosgrove-etal:87,Chynoweth-etal:1991,Okasio-etal:03,Sintes-etal:00,Kaznessis-etal:02} 
that are hardly accessible with, otherwise powerful, experimental methods, including atomic force microscopy (AFM),~\cite{We-Lee:09} 
electron microscopy (EM),~\cite{Tamm-etal:01} ellipsometry,~\cite{Ringstad-etal:08} X-ray diffraction~\cite{Seul-etal:83,Wu-etal:05}, 
vibrational spectroscopy~\cite{Kim-etal:10} etc. 
In this case computer modelling is especially useful because most often the aforementioned experimental techniques are virtually 
incompatible with the conditions under which the phenomena in question occur, e.g. in voltammetry.

In order to reckon with the difficulties in theoretical understanding and analysis of electrochemical measurements, 
below we give a brief overview of the problem and the available up-to-date analytical methods and theories. 
When the electrode surface is charged with density $\sigma$, it produces constant electric field, $E \sim \sigma$, 
which is equal to that in the interior of the capacitor with surface charge density $\pm\sigma/2$ 
on each of its plates, respectively. However, when an electrolyte solution is present, a non-trivial charge 
distribution in the vicinity of the electrode alters the electric field sensed at distance $z$ from its 
surface. As a result, the force acting on a probe charge away from the electrode decays with $z$ exponentially 
and approaches zero in the bulk solution. At the surface and in its direct proximity of a few {\AA}ngstr\"{o}m, 
though, a well-known electric double layer (EDL) phenomenon emerges, being manifested in accumulation 
of the electrode counter-ions and, respectively, depletion of the co-ions. That is, two distinct layers of opposite charge 
are observed, being arranged in an effectively neutral interface analogous to a (diffused) capacitor. 
There exists an approximate analytical (mean-field) solution of the Poisson-Boltzmann equation in this case, 
providing rather reliable expressions for the charge distribution, $\rho(z)$, and the corresponding electrostatic 
potential, $\phi(z)$, near a charged surface.~\cite{Colloidal-Domain:94} On the one hand, this solution is 
a corner-stone of the Gouy-Chapman theory~\cite{Gouy:1910,Chapman:1913} which is applicable in the cases of 
relatively weak surface potentials ($\phi(0)\apprle 0.25~V$) and in the presence of 1:1 salt in low to moderate 
concentrations ($c_{\rm salt} \apprle 0.1 M$). On the other hand, such a mean-field treatment also constitutes 
an essential part of the long-celebrated DLVO theory describing the interaction between two like-charged surfaces 
(colloids) immersed in an electrolyte solution.~\cite{Derjaguin:41,Verwey:48}

\begin{figure}
\centering
\hskip 0.0cm
\hbox{
\includegraphics[width=10.cm]{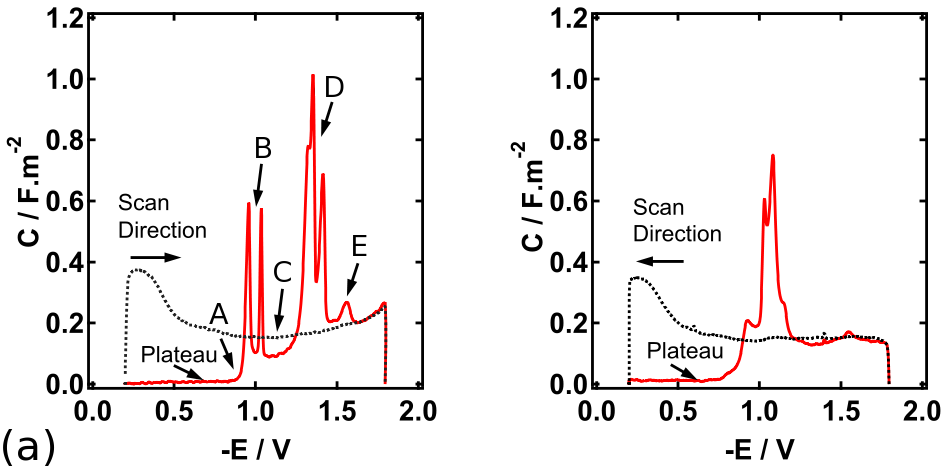}
}
\vskip 0.3cm
\hskip 0.0cm
\hbox{
\includegraphics[width=10.cm]{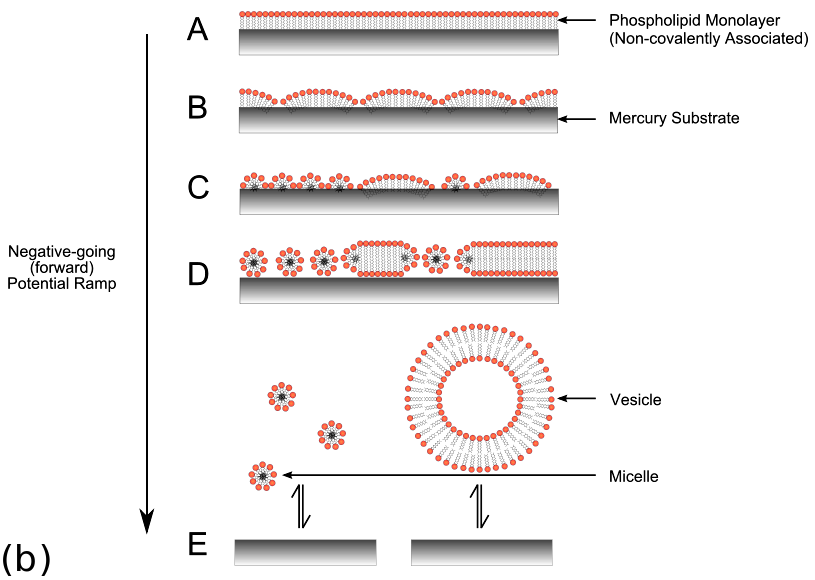}
}
\caption{\small 
Typical capacitance vs applied voltage profiles, $C(V)$, obtained in rapid cyclic voltammetry (RCV) 
experiments. The forward ($\rightarrow$) and backward ($\leftarrow$) scans are presented separately. The black dotted lines 
correspond to scans obtained for a bare electrode exposed to an 1:1 electrolyte ($0.1$~M KCl), and the solid red 
lines are obtained with a supported dioleylphosphatidylcholine (DOPC) film. 
(b) Illustration of the possible phase transitions occuring in the layered lipid films upon varying 
the applied electric potential or, equivalently, the surface charge density. The letters refer to different 
stages in between the peaks in panel a.
}
\label{Zacs_pic}
\end{figure}

Nonetheless, the picture becomes more involved when other ingredients, apart from simple ionic species, are added 
into the solution. In the case of a lipid film adsorbed onto the electrode surface, the complexity and diversity 
of the possible layered lipid structures makes the above theories hardly applicable. Therefore, more elaborate and 
accurate approaches need to be applied in order to account for the interplay between short-range (van der Waals) 
and long-range (electrostatic) interactions within the film interface. One of the main obstacles being the inhomogeneity 
of the interface in dielectric properties and local density (phase), both varying unpredictably upon charging up the electrode. 
The underlying molecular interactions are also altered, partly owing to a gradual change from hydrophobic to hydrophilic 
behaviour of the electrode, but also due to a strong, yet unknown, response in ion distribution which is especially 
difficult to account for within the interface. In particular, a well-known feature of supported monolayers exposed 
to an aqueous solution and subjected to cyclically varying electric potential, is a characteristic non-linear dependence 
of the interfacial capacitance $(C)$ upon the applied voltage $(V)$. The non-linearity in the $C(V)$ profile is normally 
evinced in one or more peaks which can be attributed to a relatively abrupt redistribution of charges and/or 
polar molecules within the monolayer interface with the solution.~\cite{LAN-FAML:90,Nelson:91,Bizzotto-Nelson:98} 
While the peculiar $C(V)$ shape appears to be specific for the combination of the monolayer properties and the solution 
constituents, the charge redistribution is thought to be associated with certain transformations in the monolayer 
structure, which can be understood as a sequence of phase transitions, as is visualised 
in Fig.~\ref{Zacs_pic}.~\cite{FAML-LAN:90,Bizzotto-Nelson:98,Conway-edts:99,Coldrick-PhD:09}
There exist also plausible theoretical (phenomenological) approaches to the problem.~\cite{Monne-etal-Nelson:03,Monne-etal-Nelson:07,Limon:08,Becucci-Guidelli:09} 
It remains, however, largely undetermined what type of transition is reflected by each peak in the $C(V)$ profile,~\cite{Coldrick-PhD:09} 
because such phenomenological hypotheses based on indirect observations cannot be easily verified in a supporting experiment 
that could reveal the resulting molecular (re-)arrangements. A better understanding and theoretical description in such cases requires 
further research, which can also be based on modelling with the aid of numerical methods. For instance, this way substantial 
insights into possible phase phenomena in the supported phospholipid layers have been gained in the earlier study~\cite{FAML-LAN:90,LAN-FAML:90} 
where self-consistent field calculations were reported for phosphatidylcholine (PC) films adsorbed from solutions of various PC concentrations. 

In this report we study by numerical methods the effect of salt on the adsorption and structure of a lipid 
film in the proximity of a gradually charged up electrode. We start with the model system previously studied 
by Leermakers and Nelson~\cite{FAML-LAN:90} and perform self-consistent field (SCF) calculations where 
the electrostatic effects are explicitly accounted for. After assessing on the mean-field level the role of the surface 
counterions (salt cations) in displacing the lipid monolayer from the surface, we address the problem by means of 
Monte Carlo (MC) simulation for a coarse-grained lipid (or surfactant) model in the presence of salt modelled explicitly 
within the restricted primitive electrolyte framework. 

\vfill
\newpage

\section{Methodology}

In what follows we employ both self-consistent field calculations and Monte-Carlo simulations.
First, within the mean-field approximation, we study the major effect of how the presence of 1:1 
salt influences the attachment of a lipid monolayer to an electrode with varying surface charge 
density. This brief SCF study is done for the same lipid model as that used previously by 
Leermakers and Nelson.~\cite{FAML-LAN:90} Second, from Monte Carlo simulations, 
we obtain a more detailed picture of the structural rearrangements occuring in the monolayer 
upon gradually increasing the surface charge, which mimics a stepwise raise of the electric 
potential applied to the electrode. In this case, in order to minimise the computational demand, 
we opted to use a coarse-grained lipid model with a single tail being of the same thickness 
as the head monomer(s). 

\begin{figure}
\centering
\hskip 0.0cm
\hbox{
\includegraphics[width=8.0cm]{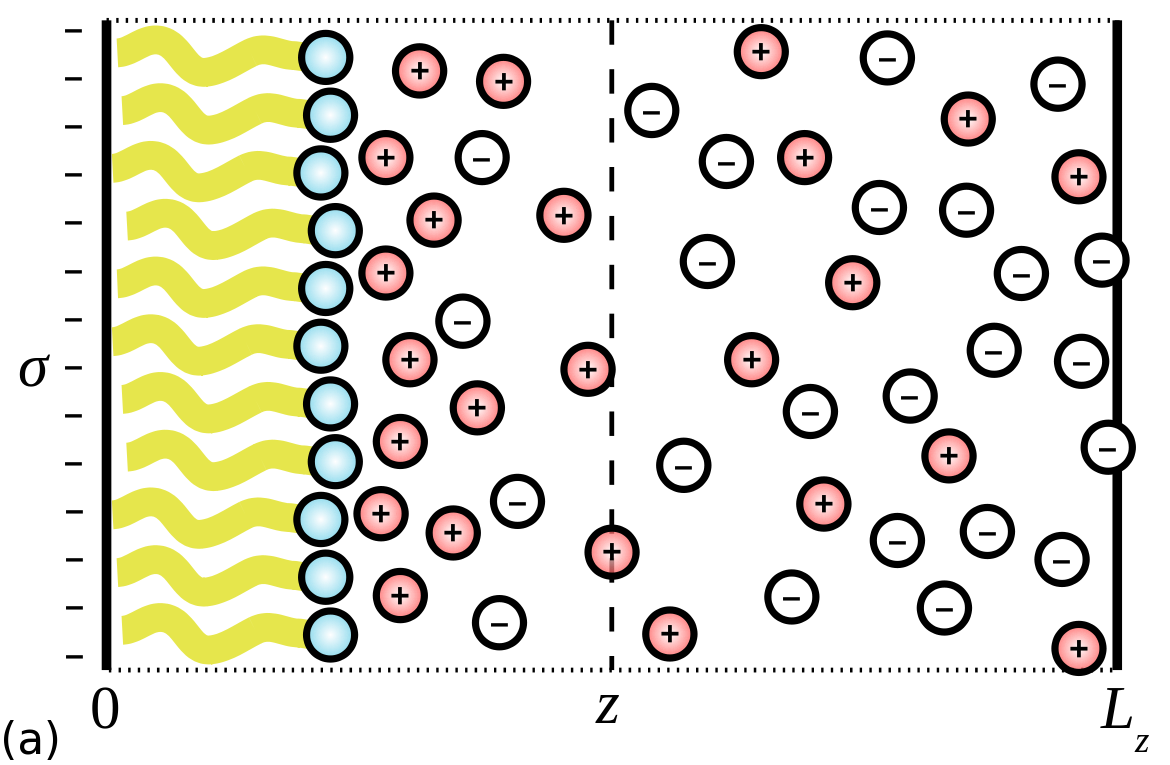}
}
\vskip 0.3cm
\hskip 0.0cm
\hbox{
\includegraphics[width=3.0cm]{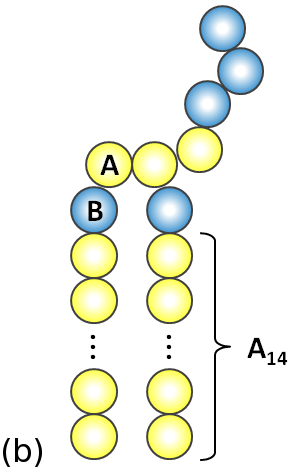}
\hskip 0.20cm
\includegraphics[width=6.0cm]{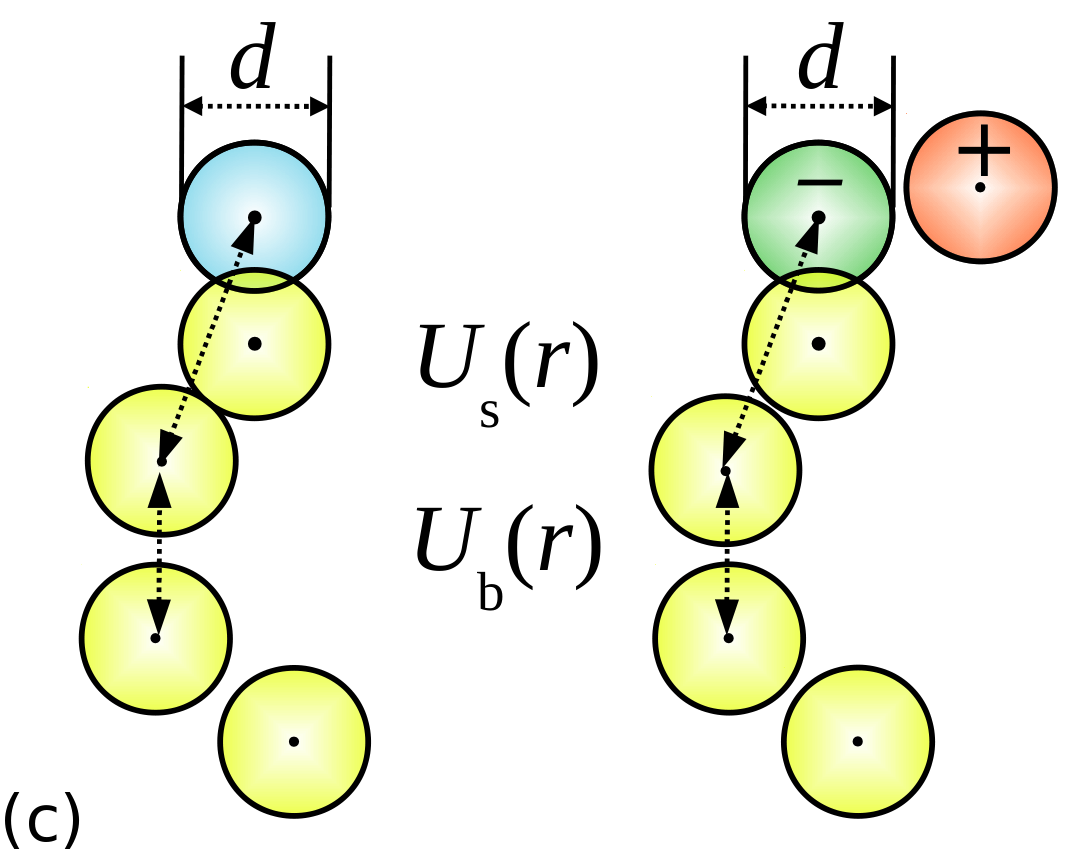}
}
\caption{\small 
(a) Schematics of the general model system. The picture illustrates the primary box used in Monte Carlo simulations. 
Within the SCF framework the picture corresponds only to the left half of the system considered, the right half being a mirror image 
of the shown part. (b) Lipid model consisting of hydrophobic A-monomers (yellow) and hydrophilic B-monomers (blue) used in SCF calculations. 
(c) Lipid models used in Monte Carlo simulations: hydrophobic beads are depicted in yellow, inert (slightly polar) head bead -- blue, 
charged beads in a zwitterionic lipid head -- green ($-e$) and orange ($+e$). See text for details.
}
\label{fig_system}
\end{figure}

The general model system in both cases is the same, as sketched in Fig.~\ref{fig_system}(a). 
The electrode surface is considered atomically smooth, as is the case with mercury at room temperature, 
and modelled by an infinite plane at $z=0$ with the surface charge distributed uniformly (when electric 
field is applied). The major difference between SCF calculations and Monte Carlo treatment is as follows. 
The mean-field problem is formulated and solved for interrelated potential and concentration (volume fraction) 
profiles defined on a one-dimentional grid along the $z$-axis under the assumption of uniform distribution 
of species within each planar ($xy$) layer corresponding to a $z$-bin. Therefore, the resulting self-consistent 
solution gives only the average density profile for each component in the system, whereas any inhomogeneities 
in the lateral directions, due to specific configuration arrangements and inter-species correlations, remain 
inaccessible. In contrast, MC simulations are performed with finite number of lipid molecules and salt ion pairs 
in a sufficiently large simulation box that is periodically repeated in the $x$- and $y$-directions. Thus, 
the correlations and inhomogeneities are explicitly included into the consideration. Here the advantage is 
that MC results are statistically exact within the model used. 

%\newpage

\subsection{Self-consistent field calculations}
\label{SCF_method}

The lipid model we use in the mean-field calculations is sketched in Fig.~\ref{fig_system}(b). 
The SCF method is similar to, albeit not the same as, that employed earlier for other lipid systems.~\cite{FAML-LAN:90,Leermakers:88}
We note that in that work the electrostatic effects were accounted for only implicitly, i.e. via the variation 
of the Flory-Huggins interaction parameters controlling the interplay between the affinities to the surface of lipid head 
groups and solvent species. In the current study we make the model more realistic by directly including 
the electrostatic effects due to the varying surface charge density and the presence of 1:1 salt, which resemble 
the actual experimental conditions.

The self-consistent field lattice theory has been initially developed by Scheutjens and Fleer~\cite{Scheutjens:85,Fleer:93} 
and later generalized to polyelectrolytes by B\"{o}hmer et~al~\cite{Bohmer-etal:90} and  Israels et~al~\cite{Israels-etal:93,Israels-etal:94}. 
The detailed description of the current implementation of the method can be found in a recent work where it has also been adapted 
to protein-polysaccharide systems.~\cite{Akinshina:08} Therefore, below we present only the main aspects of the model system 
and the method. 

The system comprises lipid molecules, salt ions, and solvent species (water) distributed between two parallel plates 
that are placed sufficiently far away from each other, $z_{\max}=300$~\AA,\ so as to minimise the interaction between them. 
The space between the surfaces is divided into planar layers, $k_z = 1, 2, 3,\dots$, parallel to the plates, 
each layer being subdivided into hexagonal lattice cells of equal volume with the lattice spacing of $a_{\rm c} = 3$~\AA. 
Every lattice cell is assumed to be fully occupied by either a lipid monomer, an ion or a water molecule, so that
the total volume fraction in each layer equals 1, i.e. $\sum\phi_\alpha(z) = 1$, where $\phi_\alpha(z)$ is the local 
volume fraction for component $\alpha$. The Bragg-Williams approximation of random mixing is applied and, thus, 
all cells in the same layer are, on average, equivalent in terms of concentration of each species (composition). 

The main assumption in the mean-field approach is that the total action on each species $\alpha$ induced by all the components 
in the system (including the same type) can be approximated by the corresponding potential of mean force, 
$u_\alpha(z;\phi_\alpha(z))$.
Thus, the two primary functions to be obtained for each species in the SCF calculation are: the average volume fraction, 
$\phi_\alpha(z)$, and the mean-field potential, $u_\alpha(z)$. Generally, these are coupled via a set of non-linear 
(differential for electrostatic contribution) equations which is solved self-consistently by iteration, provided the appropriate 
boundary conditions such as the bulk values, $\phi_\alpha^{\rm(bulk)}(z)$ and $u_\alpha^{\rm(bulk)}(z)$. In the course of the numerical 
iterative procedure,~\cite{Akinshina:08} the convergence is achieved when the profiles, $\phi_\alpha(z)$ and $u_\alpha(z)$, become 
consistent with each other, i.e. stop varying notably between consecutive iterations. 

In the current model there are five types of molecular components: the solvent -- water (denoted by index $\alpha=$ W), 
the lipid groups ($\alpha = $ A, B, where A stands for hydrophobic back-bone monomers and B -- for polar groups), cations ($\alpha = $ C), 
and anions ($\alpha = $ D). The lipid chain has two hydrophobic tails of $l_{\rm t}=14$ A-monomers and the polar head region is modelled 
by a combination of both A- and B-monomers, as shown in Fig.~\ref{fig_system}(b). 

The parametrisation is done so as (i) to mimic the system originally studied by Leermakers and Nelson,~\cite{FAML-LAN:90} 
and (ii) to investigate the pure electrostatic effect on the lipid adsorption due to the addition of surface charge and salt explicitly. 
Therefore, we study the system where the supported monolayer is initially, i.e. prior to application of the electic field, thermodynamically 
stable owing to strong hydrophobically driven adsorption of lipids onto the electrode. Below we give the Flory-Huggins parameters that 
correspond to this regime. We emphasise that the parameter values {\it are not altered} upon charging up the surface, even though 
the hydrophobic effects at the electrode would diminish under the experimental conditions. 
The hydrophobic tail monomers on lipids (type A) strongly repel both the solvent (W), $\chi_{\rm AW} = 1.6\ k_{\rm B}T$, and the polar heads (B), 
$\chi_{\rm AB} = 1.5\ k_{\rm B}T$, while the polar monomers (B) weakly attract solvent molecules, $\chi_{\rm AB} = -0.3\ k_{\rm B}T$. 
The lipid tails have no interaction with the surface (S), $\chi_{\rm AS} = 0$, while the heads and the water molecules are repelled from it, 
$\chi_{\rm BS} = 0.8\ k_{\rm B}T$ and $\chi_{\rm WS} = 4.8\ k_{\rm B}T$. The interactions of water with the surface are set to be strongly 
repulsive in order to model hydrophobic surface and to focus on the effect of {\it the competitive adsorption of lipids and ions}. The short-range 
interactions for ions (C, D) with each other and with all the other components are set to zero, $\chi_{\rm C} = \chi_{\rm D} = 0$, and ions 
carry charges, $q_{\rm C} = +e$, $q_{\rm D} = -e$, respectively. The surface has a fractional charge per lattice site, varied gradually, 
$q_{\rm S} = -0.1, \dots, -1.0\ e/9$\AA$^2$. The bulk concentration (volume fraction) of the lipids is set to $\phi_{\rm AB} = 7\ 10^{-7}$ 
and the concentration of salt is $\phi_{\rm salt} = 10^{-2}$, which approximately corresponds to the ionic strength of $0.5$~M.

\subsection{Monte Carlo simulations}
\label{MC_method}

\subsubsection{Model}
\label{MC_model}

In our simulations we use a coarse-grained lipid model and the restrictied primitive electrolyte model in a planar geometry, 
schematically depicted in Fig.~\ref{fig_system}(a). If not stated otherwise, number of lipids is $N_l=400$ and the primary 
box dimensions are $L_{xy}=L_x=L_y=80$~\AA,\ $L_z=200$~\AA, periodic boundary conditions being applied along the $x$ and $y$ 
directions. All simulations are performed with the aid of an in-house state-of-the-art FORTRAN code, based on the program used 
in previous studies.~\cite{Broukhno:00,Broukhno:01,Brukhno:09}

The lipid is represented by $n=5$ or $6$ spherical monomers (beads), with either a non-charged 
hydrophilic head monomer or a zwitterionic two-bead head group, as shown in Fig.~\ref{fig_system}(c). 
The beads are connected to form a chain with a harmonic spring potential, $U_{\rm b}(r_{i,i+1}) = 0.5\ k_{\rm f}\ r_{i,i+1}^2$, 
where $k_{\rm f}$ is the force constant and $r_{i,i+1}$ is the distance between the centers of two adjacent beads, 
$i=1,\dots, n-1$. In order to account for the stiffness within a lipid molecule we also apply a disjoining spring, 
$U_{\rm s}(r_{i,i+2}) = -0.1 k_{\rm f}r_{i,i+2}^2$, between beads with indeces $i=1,\dots, n-2$ and $i+2$. 
This is similar in spirit to how a stepwise stiffness potential is used between next-adjacent monomers in 
the hard-core model for lipids~\cite{Davis-etal:07} and the tube model for polypeptides.~\cite{Hoang-etal:04,Auer-etal:09} 
The force constant, $k_{\rm f}$, was chosen so that the mean distance, $\langle r_{i,i+1} \rangle$, varied 
in the range $4.3$ -- $4.6$~\AA.

Additionally, a truncated and shifted Lennard-Jones (LJ) potential, 
\begin{equation}
U\dr{LJ}\ur{trunc}(r)= U\dr{LJ}(r)-U\dr{LJ}(r=3d), 
\label{Uljtr}
\end{equation}
where 
\begin{equation}
\beta U\dr{LJ}(r)=-4\epsilon_{\alpha\beta}\left(\left({d\over r}\right)^6-\left({d\over r}\right)^{12}\right), 
\label{Ulj}
\end{equation}
\noindent acts between all monomer and ion pairs ($\beta = k_{\rm B}T$, $k_{\rm B}$ being the Boltzmann constant 
and the temperature $T=298$~K). In Eqs.~\ref{Uljtr} and~\ref{Ulj} $d$ is the effective bead diameter (4~\AA\ for all species) 
and $\epsilon_{\rm \alpha\beta}$ is the attractive well depth, $\alpha$ and $\beta$ being the bead types: ``t'' 
and ``h'' for lipid tail and head monomers, ``a'' and ``c'' for the salt anions and cations, respectively. 
The repulsive LJ term models monomer soft core interactions whereas the attractive contribution is responsible 
for mediating the effective short-range attraction. In particular, the hydrophobic interaction between all tail 
monomers is mimiced by using deep wells with $\epsilon_{\rm tt}=1.2,\dots, 2.0\ k_{\rm B}T$, whereas the ions and lipid 
head beads are treated as inert by using $\epsilon_{\rm aa}=\epsilon_{\rm cc}=\epsilon_{\rm hh}=0.01\ k_{\rm B}T$, 
and for all other inter-species interactions $\epsilon_{\alpha\beta}=\sqrt{\epsilon_{\alpha\alpha}\epsilon_{\beta\beta}}$, 
$\alpha \ne \beta$ (the Lorentz-Berthelot rule). 
In order to mimic hydrophobically driven adsorption of the lipid molecules onto the electrode surface, we also 
introduce an attractive rectangular well potential that is non-zero only within a short distance from the (infinite) 
planar surface, $U_{\rm ts}(z)=-\epsilon_{\rm ts}$ for $0 < z\le d_{\rm s}$, where 
$\epsilon_{\rm ts}=0,\dots, 2.0~k_{\rm B}T$ and $d_{\rm s}=2$~\AA. 

In the cases where the surface is charged and ionic species are present, the restrictied primitive electrolyte model is employed 
where the solvent is replaced by a continuous medium with the dielectric permittivity $\epsilon_r = 78.7$ (water). 
In practice this implies using the Bjerrum length, $\lb = \beta e^2/(4\pi\epsilon_0\epsilon_r)$, as a standard factor 
in all calculations concerning electrostatics; $e$ being the unit electric charge, and $\epsilon_0$ -- the dielectric constant. 
In our simulations we use $\lb = 7.126$~\AA\ corresponding to normal conditions, $T=298$~K. The total electrostatic contribution 
to the potential energy can be written as  
\begin{equation}
U_{\rm el}^{\rm tot} = U_{\rm el}^{\rm surf} + U_{\rm el}^{\rm box} + U_{\rm el}^{\rm corr}, 
\label{Uelt}
\end{equation}
\noi where the first term is due to the electric field outside of the electrode bearing the surface charge density $\sigma$ ($e/$\AA$^2$),
\begin{equation}
\beta U_{\rm el}^{\rm surf} = -2\pi \lb \sigma \sum_{i}^{N_{\rm q}} q_i z_{i}, 
\label{Uels}
\end{equation}
\noi $N_{\rm q}$ being the total number of charges in the system, including salt ions and charged head-beads on lipids if any. 
The second contribution in Eq.\ref{Uelt} is the sum of pairwise Coloumb interactions between all charged species within the simulation box, 
\begin{equation}
\beta U_{\rm el}^{\rm box} = \lb \sum_{i}^{N_{\rm q}-1} \sum_{j>i}^{N_{\rm q}} { q_i q_j \over r_{ij} },
\label{Uelb}
\end{equation}
\noi while the third term is the long-range correction to the electrostatic energy due to the ``external'' potential $\phi_{\rm ext}(z_i;\rho(z))$ 
produced by the average charge distribution outside of the primary box, $\rho(z)$,~\cite{Torrie:80,Valleau:91}
\begin{equation}
\beta U_{\rm el}^{\rm corr} = \lb \sum_{i}^{N_{\rm q}} q_i \phi_{\rm ext}(z_i;\rho(z)). 
\label{Uelc}
\end{equation}
Here $\rho(z)$ is assumed to be identical to that obtained within the box. Being dependent on the positions of all the charges in the system 
as well as the overall charge concentration profile, $\phi_{\rm ext}(z)$ is computed numerically on a grid of $n_z=400$ bins along the $z$-axis. 
Besides, one has to reiterate this calculation sufficiently often in order to account for redistribution of the charge 
along the box $z$-dimension, which is crucial for equilibration. That is, the long-range correction is calculated approximately, 
on the mean-field level. However, as long as the lateral dimensions of the box are sufficiently large, $L_{xy}>>\lb$, the approximation 
does not reduce considerably the precision of the overall calculation. 

Another important detail in our simulations is maintaining the chemical potential of the salt constant while varying the surface charge. 
Clearly, accumulation of counter-ions (and depletion of co-ions) at the interface results in notable variations of the ion concentration 
away from the interface, i.e. in the mid region of the simulation box. Therefore, one has to take special measures in order to attain 
the chemical equilibrium with the bulk solution, which is done here by use of the standard technique of random insertions and deletions 
of salt ions under the conditiom of constant chemical potential, i.e. in the grand canonical ensemble.~\cite{AllenTildesley,Frenkel-Smit:02} 
As usual, in order to keep the system electroneutral at all times, we performed insertion/deletion attempts on pairs of oppositely 
charged ions, i.e. salt species. In most cases we maintained the salt concentration about $0.3$~M in the mid region (if not stated otherwise).

\subsubsection{Simulation aspects}
\label{SCF_details}

Even though the lipid model (being similar in spirit to other models based on LJ-potentials~\cite{Chynoweth-etal:1991,Okasio-etal:03}) 
may seem oversimplified, and a handful of full-atom and coarse-grained lipid models can be found in the 
literature~\cite{Sintes-etal:00,Kaznessis-etal:02,Marrink-etal:04,Marrink-etal:07,Sansom-etal:08}, for the purposes 
of this study the chosen simplification level is justified as follows. 
Here we are not interested in resembling closely the self-assembly properties of any particular lipid molecule, neither 
do we intend to investigate the phenomena associated with the internal configurational transformations within a lipid molecule 
(such as $cis$-/$trans$-/$gauche$-isomerisation). To the contrary, we are only interested in reproducing the most common features 
of lipids and, generally, surfactants, i.e. their hydrophobically driven aggregation into stable layered structures (nanoscopic 
films) at surfaces and interfaces in contact with an aqueous solution. Therefore, prior to production runs we ascertain that 
the simulation parameters, such as the box dimensions and the strength of hydrophobic interactions, correspond to an initially 
stable monolayer film. In particular, we start all our simulations with a prearranged 
monolayer of $N_{\rm l}=400$ lipid molecules placed in the direct proximity of the (uncharged) electrode surface whereas 
the $xy$-projection of the initital configuration resembles square packing on a planar surface. Then, in order to examine 
the thermodynamic stability of the so-prepared monolayer, we perform a number of simulations in the isotension ($NP_{xy}T,~L_z ={\rm const}$) 
ensemble~\cite{Broukhno:00,Broukhno:01} for a set of $\epsilon_{\rm tt}$-values while the depth of the adsorption well is 
kept constant $\epsilon_{\rm ts}=2.0~k_{\rm B}T$. 
In this preliminary simulations we set the external lateral pressure $P_{xy}=0$ so as to find the regime of zero surface tension 
within the monolayer, i.e. the conditions at which the average $xy$-projected area per lipid does not vary, implying that the film neither 
shrinks nor expands in the lateral directions. This way we found that with $\epsilon_{\rm tt}$ in the range $1.4,\dots, 2.0\ k_{\rm B}T$ 
the monolayer is thermodynamically stable, where the observed zero-tension phase gradually transforms from the dense liquid 
(for lower values of $\epsilon_{\rm tt}$) into gel- and liquid-crystal phases (for higher values). To exemplify, the average projected 
area per lipid is varying in the range, $a_{xy}=21,\dots, 15$~\AA$^2$, which is reasonable~\cite{Colloidal-Domain:94,Larsson-Lipids:94} 
considering that the effective monomer size of 4~\AA\ is slightly underestimated, being chosen here equal to the diameter of salt ions 
-- for simplicity. In the related voltammetry 
experiments, however, the condition of zero-tension is not always fulfilled. That is, upon initial preparation of the monolayer 
by injection of a certain amount of lipids, the size (area) of the supporting it mercury droplet can be altered in 
a broad range while the number of adsorbed lipid molecules is assumed not to vary notably. Therefore, all our further (production) 
simulations, where we study the effects of the surface charge density variation, are performed with $V={\rm const}$, i.e. with 
constant area per lipid, $a_{xy}=16$~\AA$^2$ (if not stated otherwise). This setup to a good approximation resembles the experimental 
conditions where the observed peaks in $C(V)$ profiles are sharply pronounced. 

Each simulation run comprised $10^5$ passes through all the beads in the system, while either chain translation or pivot (in-chain rotation) 
move was attempted once per MC pass and two attempts of insertion/deletion of salt ion pairs complemented the pass in the cases where salt was present. 
In the same manner, in the isotension ensemble variation of the box $xy$-area was applied via an attempted change in $L_{xy}$ once per pass.

\section{Results and discussion}
\label{Results}

In the previous work of Leermakers and Nelson~\cite{FAML-LAN:90} they clearly showed that suppported lipid monolayers have a rich phase 
behaviour and are capable of phase separation within the adsorbed film, given the appropriate conditions. That is, when the lipid concentration 
in the bulk overcomes a critical threshold value (equivalent to the critical micellar concentration -- CMC, in the case of surfactants), 
depending on the strength of the effective water-surface and lipid-surface interactions, at least two possible phase transitions can 
be captured on the mean-field level. 

In the first case, presumably corresponding to a low surface charge density and, hence, somewhat reduced hydrophobicity of the surface, 
lipid tails have higher affinity to the surface than water, whereas lipid head groups partly displace tail monomers from the surface. 
This process leads to emmergence of an inhomogeneous supported film where the inhomogeneities are argued to be associated with regions 
of low lipid content, i.e. small vacancies and pores. This should be expected as flipping of a considerable fraction of head groups 
towards the surface should result in formation of bilayer patches that would occupy on average smaller surface area than the original 
monolayer with the same number of lipid molecules. As a matter of fact, a similar phenomenon 
is observed in our preliminary MC simulations under the condition of zero surface tension where $\epsilon_{\rm tt}<1.5~k_{\rm B}T$. 
In this case, however, the process of lipid flipping is driven entropically, as the tail-tail hydrophobic coupling appears to be 
insufficient to maintain the monolayer intact. The emmerging liquid bilayer merely shrinks laterally which results in a smaller 
area per lipid, cf. $\langle a_{xy}\rangle = 15.6$ and $19.6$~\AA$^2$\ as obtained with $\epsilon_{\rm tt} = 1.4$ and $1.6~k_{\rm B}T$, 
respectively. It is worth noting that only few lipid molecules escape from the adsorbed film in the process of this transformation, 
because in this simulations the hydrophobic interaction of tail monomers with the surface is still kept strong, $\epsilon_{\rm ts} = 2.0~k_{\rm B}T$. 

In the second scenario, when the water affinity to the surface is set higher than that of lipid tails, i.e. upon further implicitly 
increasing the surface charge which results in a gradual switch from hydrophobic to hydrophilic surface, the lipids can be eventually 
replaced in the immediate proximity of the surface by a layer of water. Again, the authors argue that the outcome in this case is 
an inhomogeneous planar phase with alternating regions of adsobed and desorbed bilayer. 
Although within the mean-field framework 
direct observation of the indicated phase separation and transitions is not possible, the insights into the supported lipid phase 
behaviour gained in this SCF study are intriguing. Indeed, as we will see below, the structural transformations within adsorbed 
surfactant (or lipid) films to a large extent resemble the picture outlined above, whereas the underlying molecular mechanisms are 
found to be governed by strong electrostatic effects owing to redistribution of charged species within the interface upon variation 
of the electrode charge density.

\subsection{SCF: counterions displace lipids from the electrode}
\label{Results_SCF}

As was mentioned above, we start with the SCF calculations for the same lipid (DOPC) model as that investigated by Leermakers 
and Nelson.~\cite{FAML-LAN:90} The major difference being that here we explicitly include 1:1 electrolyte and gradually charge-up 
the surface, while all the short-range Flory-Huggins parameters are kept constant, as described in section~\ref{SCF_method}. 
We recapitulate that this way the electrode surface is maintained strongly hydrophobic at all times, 
and only pure electrostatic effects on the monolayer adsorption are studied. 

\begin{figure}
\centering
\hskip 0.00 cm
\hbox{
\includegraphics[width=8.0cm]{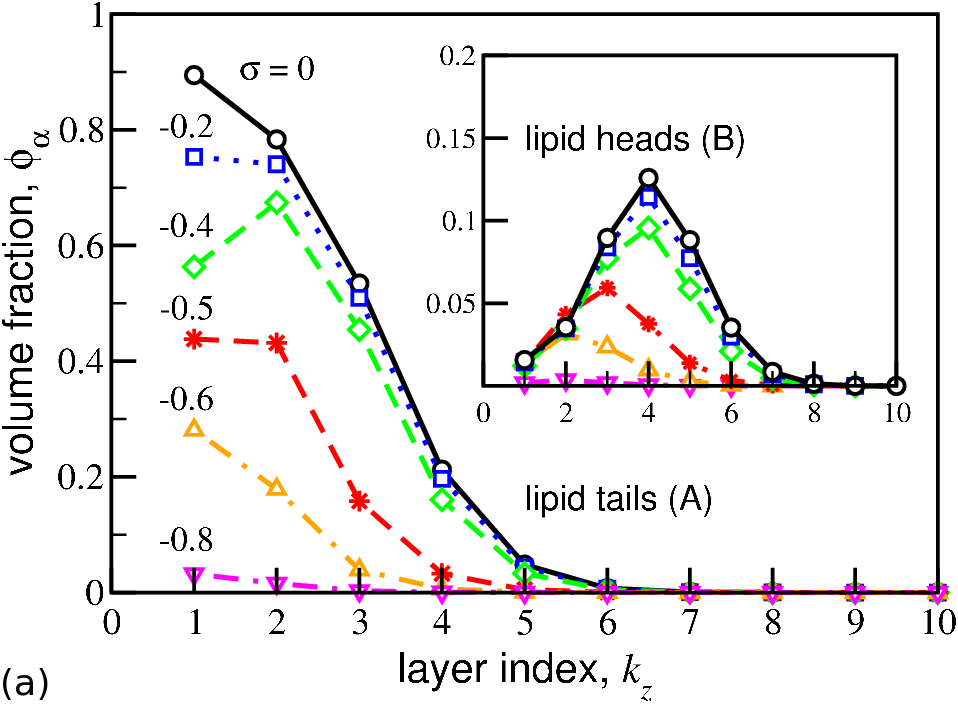}
}
\vskip 0.5 cm
\hskip 0.00 cm
\hbox{
\includegraphics[width=8.0cm]{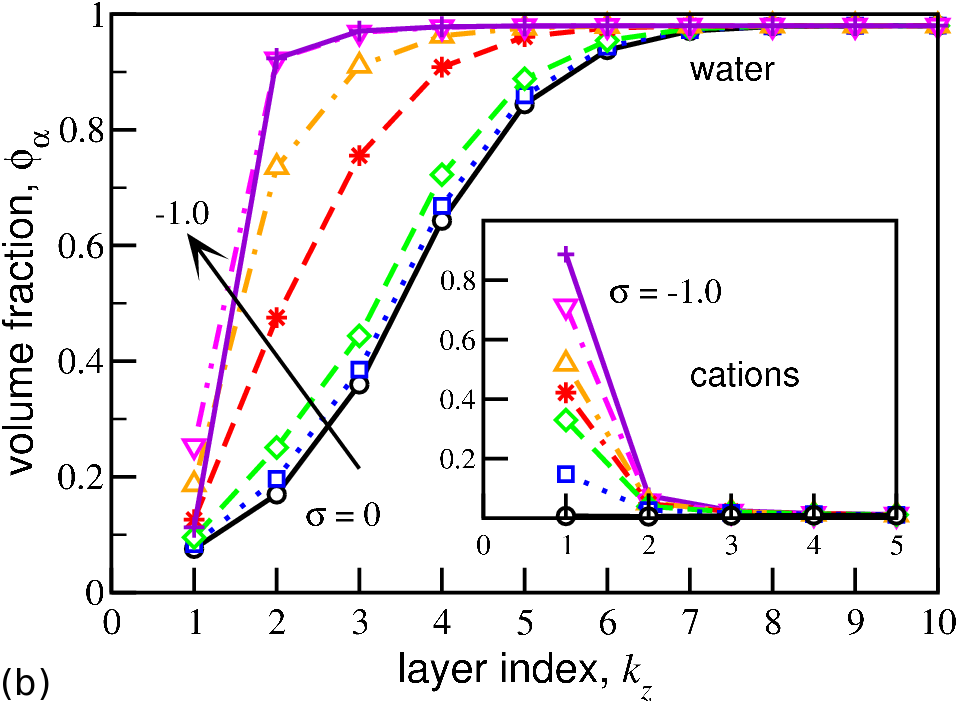}
}
\vskip 0.5 cm
\hskip 0.00 cm
\hbox{
\includegraphics[width=8.0cm]{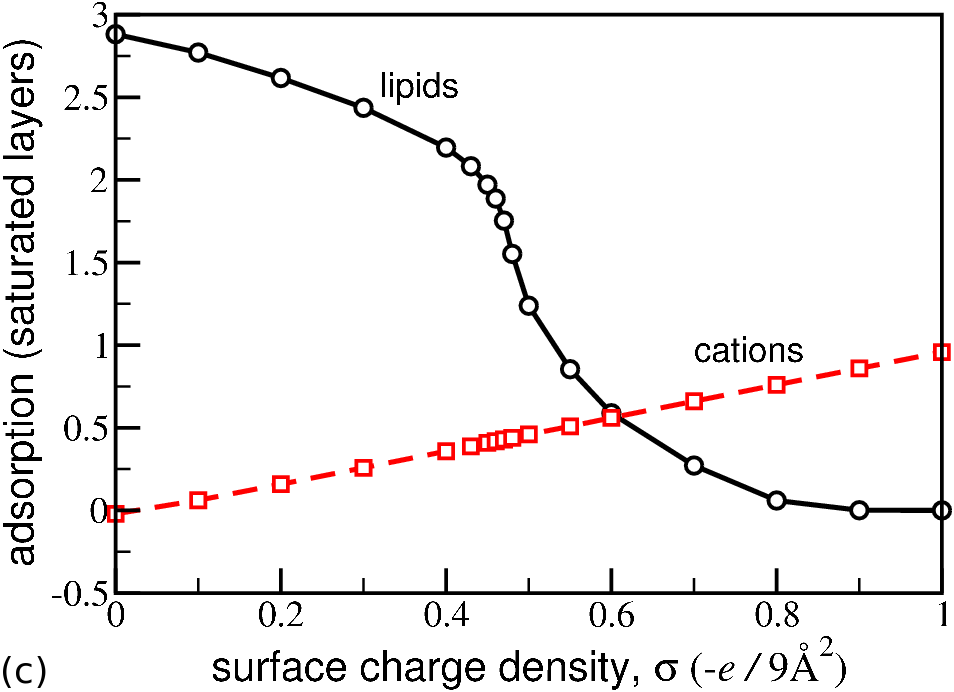}
}
\caption{\small 
(a) Variation of volume fraction profiles for lipid monomers near a gradually charged up hydrophobic surface located at $z=0$. 
The surface charge density is raised stepwise ($e/\AA^2$): $\sigma = 0$ (black lines, circles), $-0.02$ (dotted, squares), $-0.04$ (dashed, diamonds), 
$-0.06$ (dashed, stars), $-0.08$ (dot-dashed, triangles-up), $-0.1$ (dot-dashed, triangles-down). 
(b) Same as in panel a, but for surface counterions (cations) and water.
(c) Adsorption isotherms for lipids (solid, circles), and cations (dashed, squares). 
}
\label{Density_SCF}
\end{figure}

The obtained volume fraction distributions for lipid monomers, cations (electrode counterions), and water are presented 
in Fig.~\ref{Density_SCF}. Also shown are the adsorption isotherms for lipids and cations,~Fig.\ref{Density_SCF}(c). 
It is clearly seen that, as the surface charge density increases, the amount of lipids adsorbed onto the surface progressively 
decreases, whereas at the same time the counterions are gradually accumulated by the electrode, similarly to what would be 
observed in the case of a simple electrolyte (Poisson-Boltzmann solution), i.e. in the absence of the monolayer. Eventually, when 
the surface charge density approaches the value of one unit charge per lattice cell, $-e/9$~\AA$^2$, the lipid content at the surface 
becomes negligible -- the lipids completely desorb. Examination of the adsorption isotherms reveals that, as the surface charge increases, 
accumulation of cations by the electrode proceeds linearly, while the lipid desoprtion process goes in a pronounced stepwise manner, 
with the inflection point at $\sigma\approx -0.5~e/9$~\AA$^2$. Thus, our SCF calculations support the view on the interface transformation 
as a phase transition. 

We also reckon that in none of the cases water content notably exceeds that of counterions in the layer next to the charged surface. 
This is expected because in our mean-field treatment water molecules interact with the surface only via the short-range 
(nearest layer) interaction that is relatively weak in comparison to the electrostatic action experienced by a cation 
in direct contact with a negatively charged surface. 
Even if one takes into account the dipole moment of the water molecule (not included here) and considers its interacion 
with a charged site on the surface, the charge-dipole interaction decays with the distance as $r^{-2}$ whereas Coloumb interaction 
between two point charges decays only as $r^{-1}$. By the same token the electrostatic contribution to the electrode affinity of water 
is significantly lower than that for a counterion.

\subsection{MC: bilayer formation for entropic reasons}
\label{Results_MC1}

\begin{figure}
\centering
\hskip 0.00 cm
\hbox{
%(a)
\includegraphics[height=5.cm]{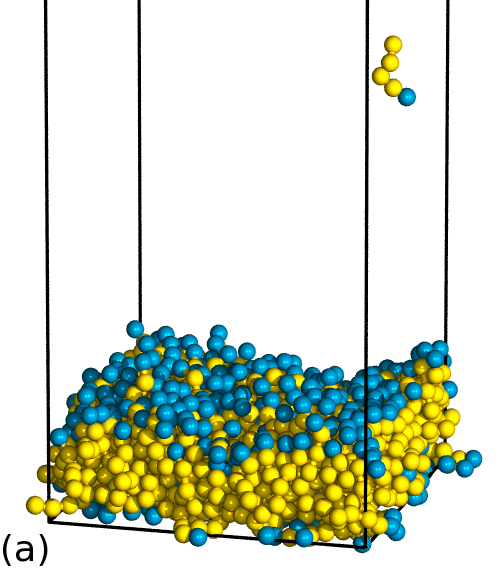}
\hskip 0.50 cm
%(b)
\includegraphics[height=5.cm]{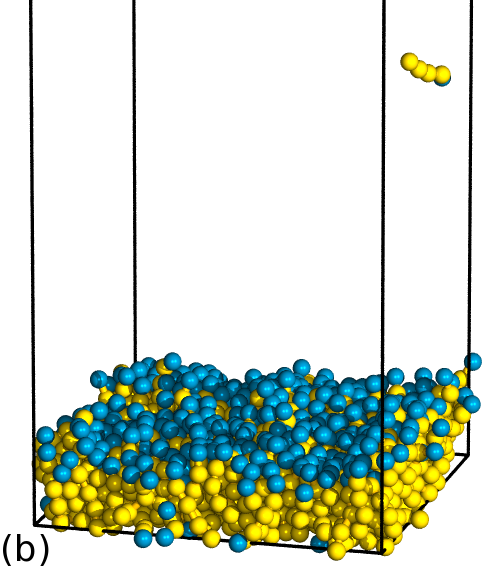}
\hskip 0.50 cm
%(c)
\includegraphics[height=5.cm]{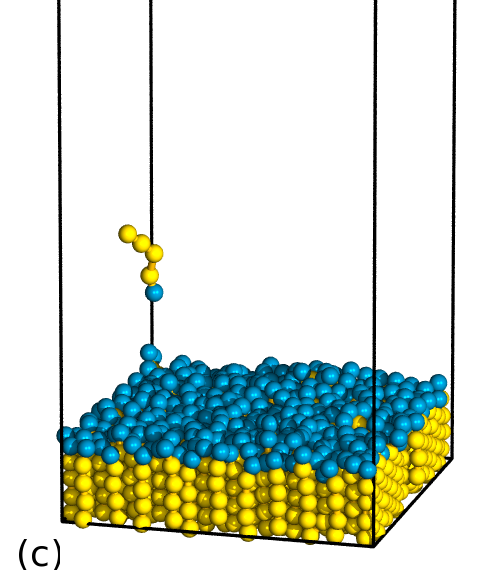}
}
\caption{\small 
Snapshots for systems with zero lateral tension, simulated in the isotension ensemble ($NP_{xy}T,~L_z={\rm const},~P_{xy}=0$). 
Lipid heads are inert, $q_{\rm h}=0,~\epsilon_{{\rm h}\alpha} =0$, whereas the depth of the tail-tail hydrophobic potential is varied ($k_{\rm B}T$): 
(a) $\epsilon_{\rm tt} = 1.4$, (b) $1.6$, (c) $2.0$. Average area per lipid, $\langle a_{xy}\rangle=15.6,~19.6,~15.4$~\AA$^2$, respectively.
}
\label{Snaps_short_neutral}
\end{figure}

\begin{figure}
\centering
\hskip 0.00 cm
\hbox{
%(a)
\includegraphics[width=8.0cm]{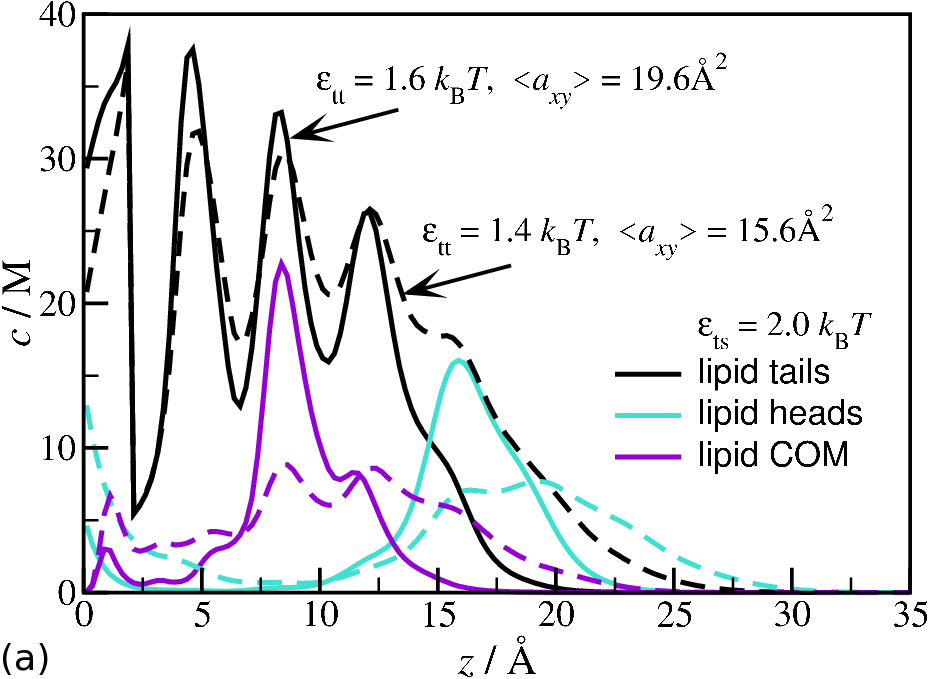}
}
\vskip 0.20 cm
\hskip 0.00 cm
\hbox{
%(b)
\includegraphics[width=8.0cm]{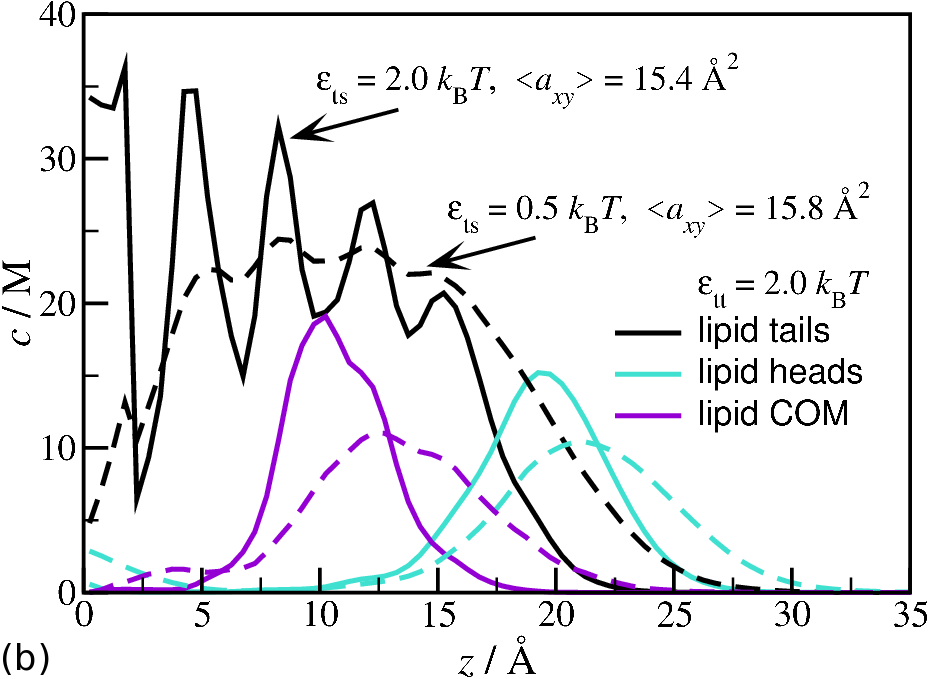}
}
\caption{\small 
Concentration profiles for lipids composed of one inert head and four hydrophobic tail beads next to a uncharged hydrophobic surface. 
The colour code in both panels: black lines -- lipid tails, blue -- lipid heads, and violet -- lipid COM:s.
(a) Solid lines correspond to $\epsilon_{\rm tt} = 1.6$ and dashed to $\epsilon_{\rm tt} = 1.4$, while 
the depth of the tail-surface interaction is constant, $\epsilon_{\rm ts} = 2.0~k_{\rm B}T$. 
(b) Solid lines correspond to $\epsilon_{\rm ts} = 2.0$ and dashed to $\epsilon_{\rm ts} = 0.5$, while 
the strength of the tail-tail interaction is constant, $\epsilon_{\rm tt} = 2.0~k_{\rm B}T$. 
See text for details. 
}
\label{Density_short_neutral}
\end{figure}

In the light of the mean-field results of Leermakers and Nelson~\cite{FAML-LAN:90} we briefly assess the effect of variation 
of hydrophobic tail-tail and tail-surface interactions in MC simulations with neutral surface, i.e. corresponding to the electrode 
kept at the potential of zero charge (PZC). In Fig.~\ref{Density_short_neutral} the lipid concentration profiles obtained under 
the conditions of zero-tension are given for a set of different combinations of $\epsilon_{\rm tt}$ and $\epsilon_{\rm ts}$ values. 
Even though in all cases the lipid film remains at least partially attached to the surface, we observe a strong tendency to formation 
of a bilayer when either of the parameters mediating hydrophobic interactions is lowered. As is also supported by the simulaton snapshots 
in Fig.~\ref{Snaps_short_neutral}, this tendency is especially pronounced in the cases where $\epsilon_{\rm tt}<\epsilon_{\rm ts}=2.0~k_{\rm B}T$, 
i.e. with strongly adsorbing liquid film, Fig.~\ref{Density_short_neutral}(a), whereas in the gel phase (b), 
$\epsilon_{\rm tt}=2.0$, a considerable drop in the tail-to-surface affinity, $\epsilon_{\rm ts}=2.0\to 0.5$, results mainly 
in partial desorption of the lipid film with only a small fraction of lipid molecules flipped towards the surface. 

Notably, in the first case where $\epsilon_{\rm tt}=1.4$ and $\epsilon_{\rm ts}=2.0$, the concentration of lipid head monomers 
at the surface is higher than that at the outer side of the bilayer. It is, though, obvious from the figure that for entropic 
reasons the lipids oriented outwards possess more freedom (entropy) in their motion, which leads to a broader distribution 
of head monomers in comparison with that near the surface. Similar trend is also seen in the distributions of centers of mass (COM)
in both cases of liquid film, where a few bumps have developed towards the outer side of the bilayer and only one distinct peak 
is located by the surface -- right within the hydrophobic potential well. These latter peaks are indicative of the configurations 
adopted by those lipid molecules that happended to turn their head groups inwards. That is, to maintain their hydrophobic contact 
with the surface these lipids oriented roughly parallel to it. Analogously, a small bump in the vicinity of the surface hydrohobic 
region is exhibited by the COM profiles in the partially desorbed gel phase, $\epsilon_{\rm tt}=2.0$ and $\epsilon_{\rm ts}=0.5$.

The observed phase behaviour is in line with the first scenario suggested by Leermakers and Nelson. For example, one can anticipate 
that under the conditions of constant amount of lipids per unit surface area the supported {\it liquid} layer should be distributed 
on the surface inhomogeneously, as was indicated by the phase diagram in Ref.~\cite{FAML-LAN:90} However, one has to remember that 
the mechanism of transformation into a bilayer in these simulations is intrinsically entropic, because lipid head monomers remain 
effectively inert, $\epsilon_{\rm hs}=0$, i.e. we do not assume any affinity to the surface for head groups. 

\subsection{MC: lipid tail -- cation race for charged surface}
\label{Results_MC2}

Below we present our MC results obtained for systems with constant area per lipid, $a_{xy}=16$~\AA$^2$, where 
a 1:1 electrolyte is explicitly included and the surface charge density is progressively raised in steps, $\sigma = -0.01,~-0.05,~-0.1~e/$\AA$^2$. 
With this procedure we mimic a few intermediate steps in the voltammetry experiments so as to explore the spectrum 
of possible phenomena occuring at different values of the applied electric potential. In these simulations we opted 
to set $\epsilon_{\rm ts}=\epsilon_{\rm tt}=2.0~k_{\rm B}T$, i.e. all hydrophobic interactions have equal strength, 
and, as we mentioned above, the lipid monolayer appears initially ($\sigma = 0$) to be in a gel phase. As in our SCF treatment above, 
while the electrode surface is charged up, we do not reduce the surface affinity of lipid tails so as to examine the electrostatic 
effects per se. 

\begin{figure}
\centering
\hskip 0.00cm
\hbox{
\includegraphics[height=5.cm]{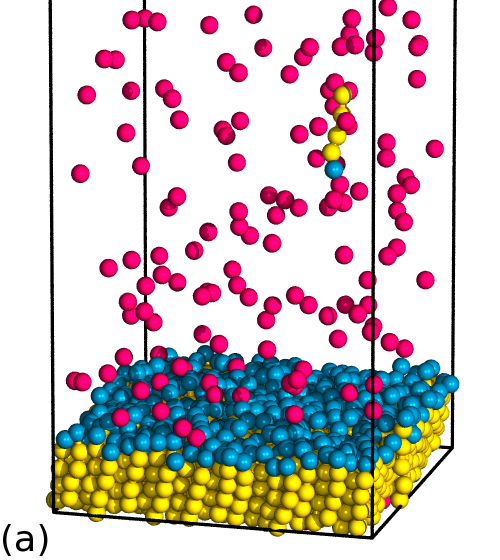}
\hskip 0.50cm
\includegraphics[height=5.cm]{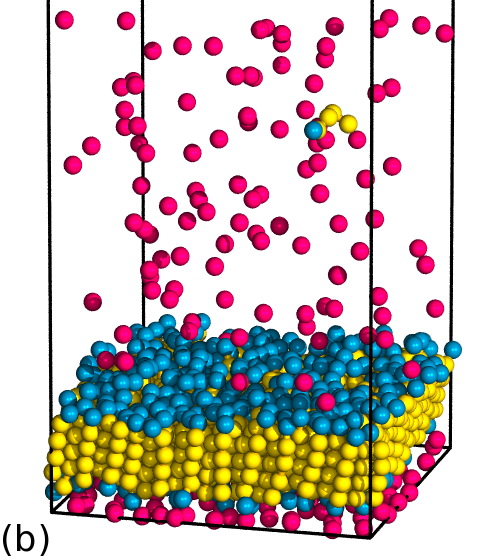}
\hskip 0.50cm
\includegraphics[height=5.cm]{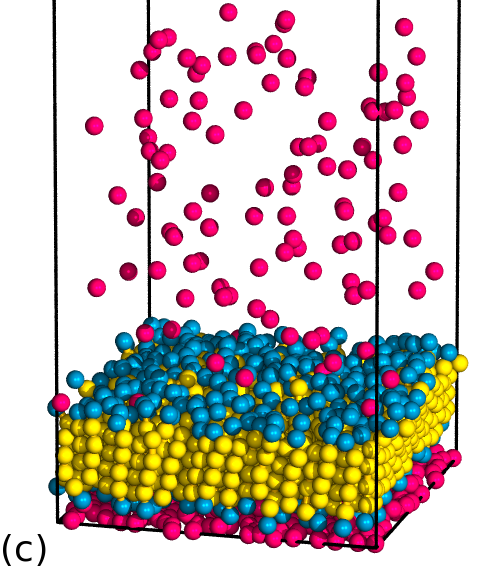}
}
\caption{\small 
Snapshots for the system with inert lipid heads, $q_{\rm h}=0,~\epsilon_{{\rm h}\alpha} =0$, simulated in $NVT$ ensemble, 
$a_{xy}=16$~\AA$^2$. The surface charge density is varied ($e/$\AA$^2$): (a) $\sigma = -0.01$, (b) $-0.05$, (c) $-0.1$.
For clarity anions are not shown.
}
\label{Snaps_short_charged}
\end{figure}

\begin{figure}
\centering
\hskip 0.00 cm
\hbox{
%(a)
\includegraphics[width=8.0cm]{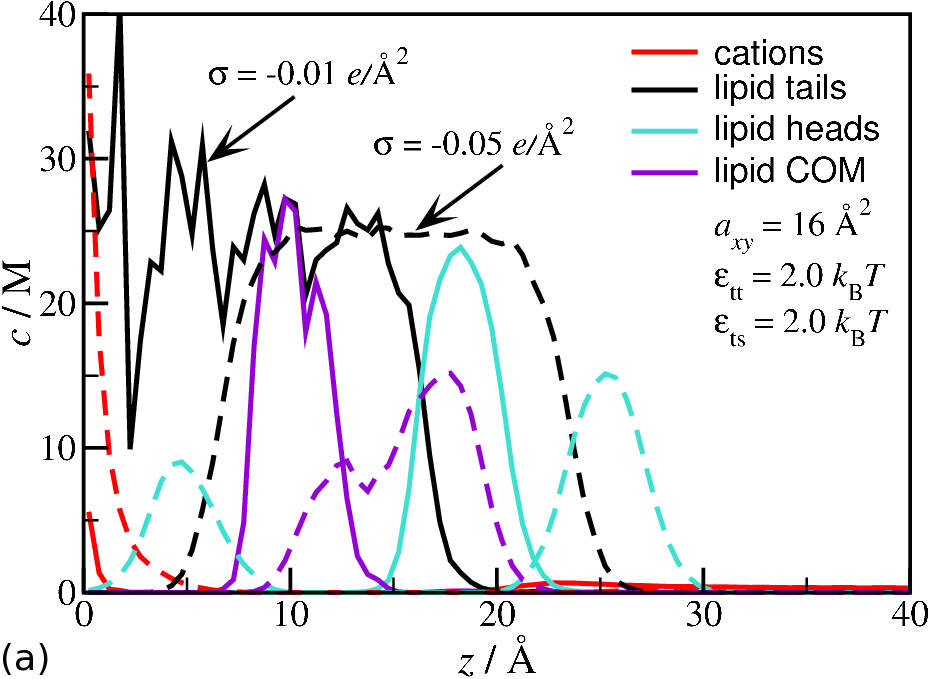}
}
\vskip 0.20 cm
\hskip 0.00 cm
\hbox{
%(b)
\includegraphics[width=8.0cm]{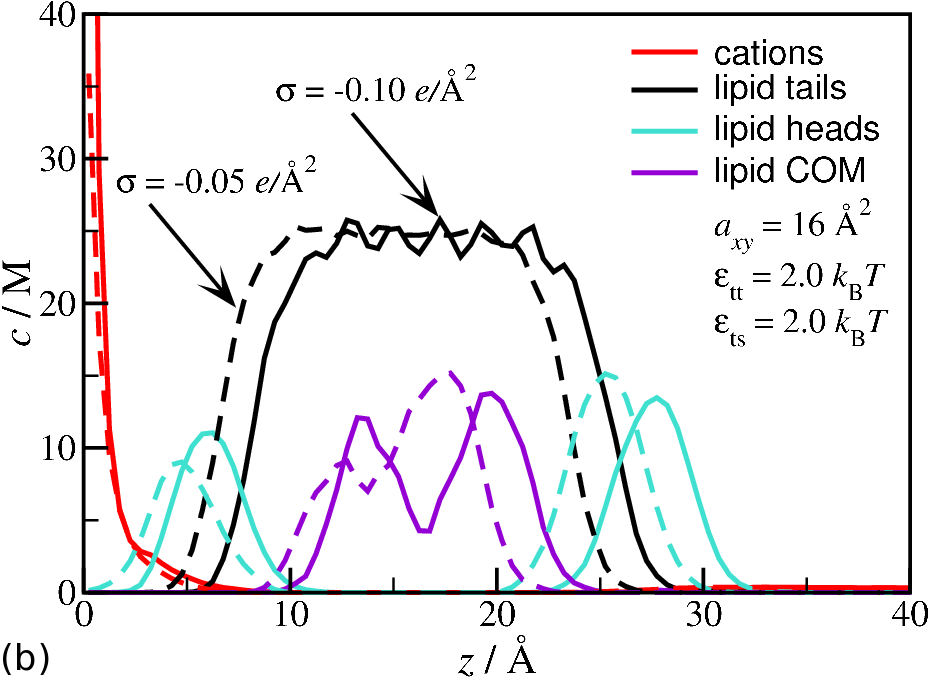}
}
\caption{\small 
Same as in Fig.~\ref{Density_short_neutral}, but for systems where the electrode surface is charged up in the presence 
of 1:1 electrolyte. The profiles shown in red are for cations (surface counterions), the colour code for lipids as above. 
(a) Solid lines correspond to $\sigma = -0.01$ and dashed to $\sigma = -0.05$.
(b) Solid lines correspond to $\sigma = -0.1$ and dashed ones are identical to those in panel a.
The area per lipid, $a_{xy}=16$~\AA$^2$, and strength of hydrophobic interactions are kept constant, 
$\epsilon_{\rm tt} = \epsilon_{\rm ts} = 2.0~k_{\rm B}T$. See text for details. 
}
\label{Density_short_charged}
\end{figure}

In this section we consider lipids with $n=5$ beads and their head monomers totally inert (effectively ``polar''), $q_{\rm h}=0$ 
and $\epsilon_{{\rm h}\alpha}=0$, where $\alpha$ denotes either the surface, lipid tail or head monomers, or salt ions. 
While Fig.~\ref{Snaps_short_charged} visually exemplifies the phenomena observed, in Fig.~\ref{Density_short_charged} the resulting 
concentration profiles are shown for three cases with stepwise increments in the surface charge. 
Similarly to what we learnt on the mean-field level in section~\ref{Results_SCF}, the lipid monolayer, being progressively 
displaced from the surface by the counterions, constantly drifts away from the surface. 
By comparing the distributions and snapshots for higher surface charges, $\sigma =-0.05,~-0.1$, we also clearly see how 
the monolayer arrangement gradually transforms into a bilayer structure. At intermediate surface charge the bilayer have 
an asymmetric distribution of lipids between its two sides, which becomes almost ideally symmetric as the surface charge 
is raised further and the lipid film departs from the electrode. In both cases the observed bilayer asymmetry is, of course, 
expected to disappear in fully equilibrated systems. Interestingly, one can anticipate the development of 
the second counterion layer next to the one accumulated directly at the electrode surface. That is, the initial electric 
field through an unperturbed monolayer is so strong (the potential drop at the surface $\phi_{z=0}\approx -2~V$), that even 
a gel-phase lipid film does not present an obstacle for the counterions, see the ``initial'' lines in Fig.~\ref{Potential_short_charged}. 
Eventually, after the electrode becomes screened by a sufficient amount of counterions, the gradient in the electrostatic potential 
is substantially reduced -- the equilibrium profiles in Fig.~\ref{Potential_short_charged}, by which time the lipids are completely removed 
from the hydrophobically favourable region by the surface. 
We emphasise that in none of the simulations with inert lipid heads we captured the displaced lipid slab moving backwards 
to the surface. This implies that finally it should end up either in a solution as a vesicle (liposome) or at the air-water 
interface as an oily monolayer film. In this respect, it is worth noting that the result of the competition for the charged 
surface between lipids and couterions is not as obvious in the real-world circumstances, because head groups of different 
lipids can be either strongly polar (as well as polarisible) or charged, which certainly should affect their behaviour 
when the electric field is applied. For example, the case of zwitterionic lipid model is considered in the next section.

\subsection{MC: monolayer desorption -- bilayer readsorption}
\label{Results_MC3}

\begin{figure}
\centering
\hskip 0.00cm
\hbox{
\includegraphics[height=5.cm]{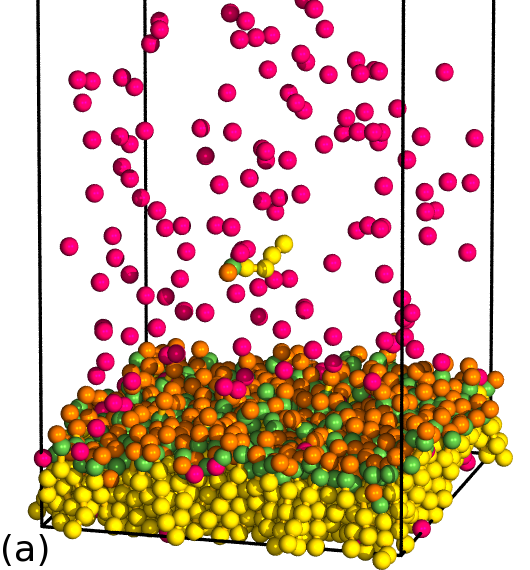}
\hskip 0.20cm
\includegraphics[height=5.cm]{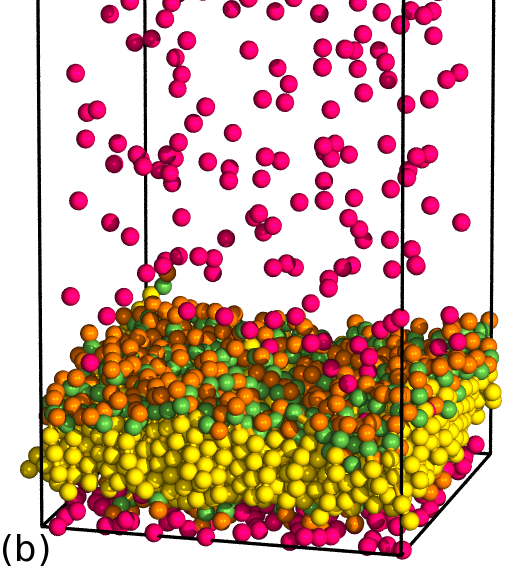}
\hskip 0.20cm
\includegraphics[height=5.cm]{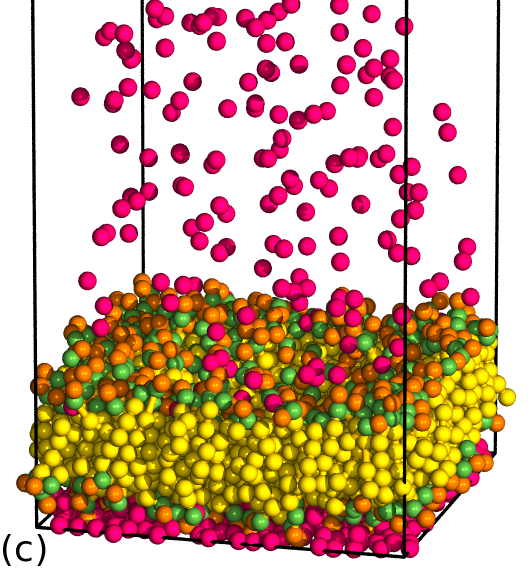}
}
\caption{\small 
Snapshots for the system with zwitterionic lipid head groups, $n=6$, $q_{\rm h}=\pm e,~\epsilon_{{\rm h}\alpha} =0$, simulated in $NVT$ ensemble, 
$a_{xy}=20.8$~\AA$^2$. The surface charge density is varied ($e/$\AA$^2$): (a) $\sigma = -0.01$, (b) $-0.05$, (c) $-0.1$.
For clarity anions are not shown.
}
\label{Snaps_short_charged2}
\end{figure}

\begin{figure}%[top]
\centering
\hskip 0.00 cm
\hbox{
%(a)
\includegraphics[width=8.0cm]{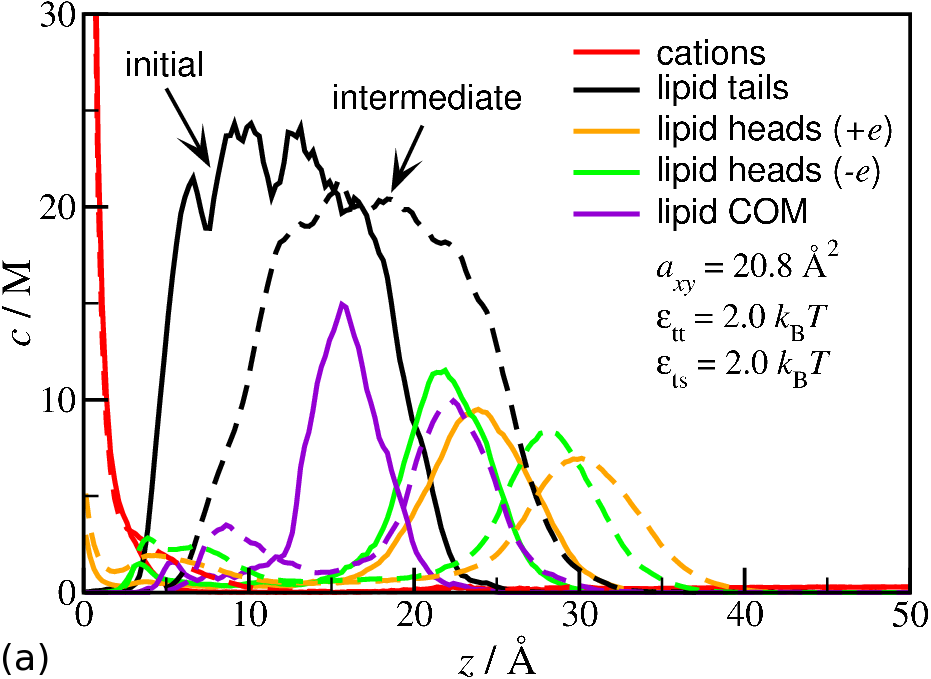}
}
\vskip 0.20 cm
\hskip 0.00 cm
\hbox{
%(b)
\includegraphics[width=8.0cm]{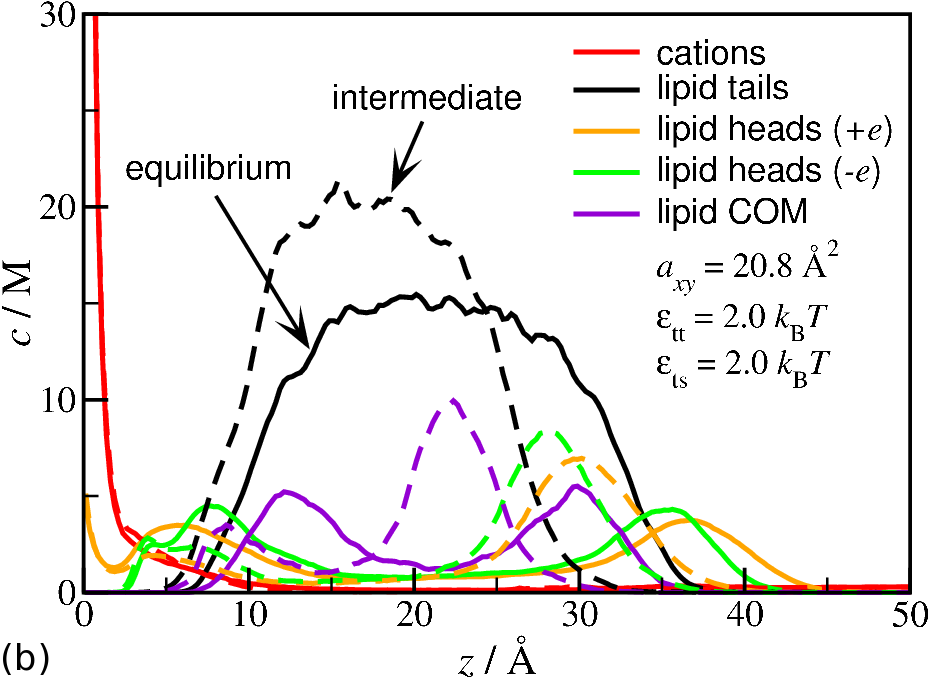}
}
\caption{\small 
Same as in Fig.~\ref{Density_short_charged}, but obtained with zwitterionic phospholipids carrying both negative (PO$_2^-$) 
and positive charges (NH$_3^+$) in the head group. The surface charge is kept constant, $\sigma = -0.1$~$e/\AA^2$. 
The colour code for cations, lipid tails and COM:s as above, whereas orange and green profiles are for positively and negatively 
charged lipid beads, respectively. 
The results from three consecutive simulations are given. 
(a) Solid lines -- initial simulation (monolayer desorption), dashed -- intermediate (bilayer formation). 
(b) Solid lines -- equilibrium (readsorption), dashed -- as in panel a.
}
\label{Density_short_charged2}
\end{figure}

Finally, we turn to lipids with zwitterionic head groups bearing both positive and negative charges. In this case $n=6$, 
implying that in the lipid model used above we charge up the head bead with negative unit charge, $-e$, and add another head bead 
with positive charge, $+e$, which correspond to phosphate, PO$_{2}^{-}$, and amine, NH$_{3}^{+}$, charged groups in phospholipids 
of lecithin type, e.g. phosphatidylcholine (PC). 
Once again we kept the same set of LJ-parameters, $\epsilon_{\rm tt} = \epsilon_{\rm ts} = 2.0~k_{\rm B}T$, and started with a preliminary 
$NP_{xy}T$ simulation where $P_{xy}=0$, which resulted in a larger average area per lipid than what was obtained with inert head beads, 
cf. $\langle a_{xy}\rangle=20.8$ and $15.4$~\AA,\ respectively. All further simulations for this system, where the surface charge was 
varied, were performed in $NVT$ ensemble with $a_{xy}=20.8$~\AA$^2$. The counterion -- lipid competition at the electrode 
with low to moderate charge density proceeded in the same fashion as we discussed above, cf. Figs~\ref{Snaps_short_charged} 
and~\ref{Snaps_short_charged2}. The desorption in this case does not necessarily lead to departure of the forming bilayer from 
the surface but, to the contrary, at sufficiently high surface charge (or applied electric field) is followed by its readsorption, 
cf. Figs~\ref{Density_short_charged} and~\ref{Density_short_charged2}, -- the process that is driven solely by the electrostatic interactions. 

Having said that, we focus on the structural phenomena observed with highly charged surface, $\sigma=-0.1$, that are illustrated 
in Fig.~\ref{Density_short_charged2}. Like in the previous cases, initially the lipid film is 
efficiently displaced from the electrode by a distance of about $5 - 10$~\AA. At this stage, in terms of MC time roughly 
corresonding to almost complete equilibration in Figs.~\ref{Density_short_neutral} and~\ref{Density_short_charged}, only 
a small fraction of lipid molecules re-oriented towards the surface, whereas the lipid film thickened and started exhibiting 
less structure, cf. the solid and dashed lines in Fig.~\ref{Density_short_charged2}(a). However, in the course of a subsequent (longer) 
simulation the number of lipids flipping their heads inwards the interface constantly increased, producing progressively growing 
bumps on the inner side of the density profiles for lipid heads and COM:s -- the dashed lines in Fig.~\ref{Density_short_charged2}. 
That is, a bilayer has formed already at this intermediate relaxation stage. Finally, the lipid film becomes {\it electrostatically 
attracted} to the double layer formed by the electrode and its counterions (a small amount of coions being present as well), 
which is apparent from a considerable increase in the amount of inwards-oriented head groups at later stages -- the solid lines in 
Fig.~\ref{Density_short_charged2}(b). At the same time the lipid film gains a well pronounced bimodal structure in its COM profiles, 
by which the transformation into a bilayer is accomplished. 

\begin{figure}
\centering
\hskip 0.00cm
\hbox{
\includegraphics[width=8.0cm]{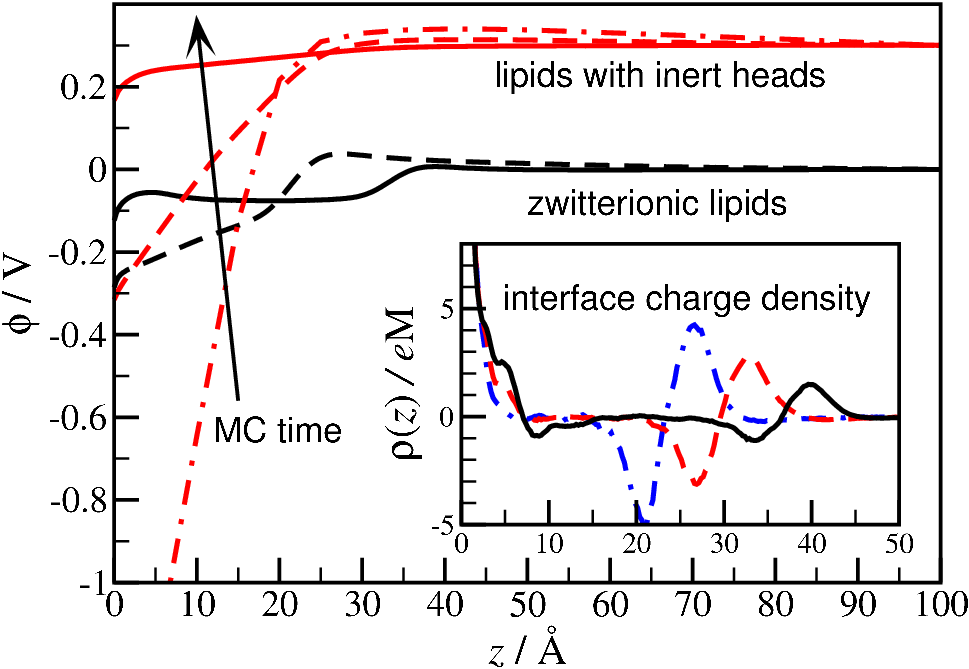}
}
\caption{\small 
Evolution of the electrostatic potential profiles, $\phi(z)$, outward a charged surface ($z=0$) for the systems of lipids 
with inert head monomers (red upper lines; shifted by $+0.3~V$ for clarity; $\phi_{z=0}\approx -2~V$) and zwitterionic head groups 
(black lower lines). Dot-dashed line -- the initial $\phi(z)$ distribution, dashed -- the intermediate ones, and 
solid -- the equilibrium profiles. 
Inset: the total charge distribution in the interface, $\rho(z)$, obtained at three stages in simulation of zwitterionic lipids: 
blue dot-dashed -- the initial charge density profile at a stage where electrode counterions started accumulating at its surface, 
red dashed -- the intermediate (non-equilibrium) charge distribution, black solid -- the resulting (equilibrium) charge profile. 
}
\label{Potential_short_charged}
\end{figure}

Obviously, at different stages of this complex process the amount of salt ions (especially cations) accumulated at the electrode constantly 
varies, which affects the properties of both the emerging electric double layer and the entire interface including the layered lipid film. 
As a result, in the case with charged lipid head groups the distribution of the electric potential through the interface is not as trivial 
as with inert head monomers. The differences between the two cases, being brought out in Fig.~\ref{Potential_short_charged}, are of course 
expected because charges in the lipid heads produce an additional contribution to the total electric field, which often results in a potential 
difference through the head group region of up to $0.25~V$.~\cite{Colloidal-Domain:94} In both cases a noticeable electrostatic barrier 
is present around the outer monolayer surface which precedes the (initially dramatic) drop in the potential. As is clear from 
Fig.~\ref{Potential_short_charged}, upon equilibration this barrier virtually disappears. Of greater interest is the fact that 
with zwitterionic lipids we observe how the charge distribution within the interface forms a multilayer of alternating charge -- 
first, starting with the emergence of an electric double layer followed by the charge reversal in the head group area 
and, eventually, equilibrating into two distinct regions of charge alternation: a ``triple'' layer ($-/+/-$) by the electrode surface 
(see inset in Fig.~\ref{Potential_short_charged}; the negatively charged surface layer is not shown)
and a double layer ($-/+$) at the outer side of the film. Without doubt, this remarkable redistribution of charge is associated with 
a substantial variation in the capacitance of the layered lipid interface, i.e. $C(V)$ peaks registered in voltammetry experiments. 

\ 

\begin{figure}
\centering
\hskip 0.00cm
\hbox{
\includegraphics[height=5.cm]{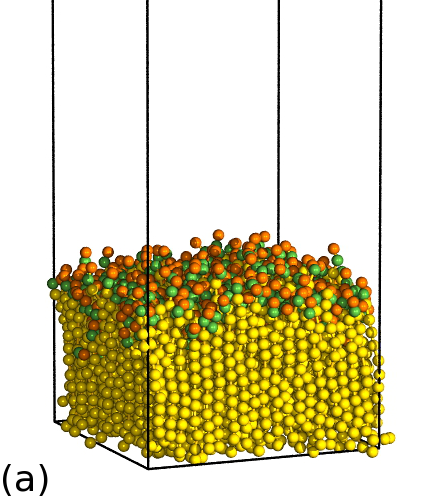}
\hskip 0.50cm
\includegraphics[height=5.cm]{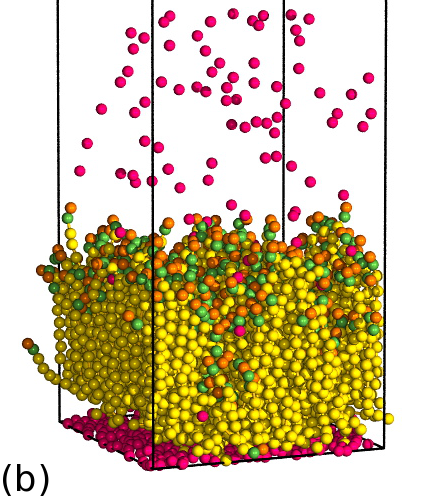}
\hskip 0.50cm
\includegraphics[height=5.cm]{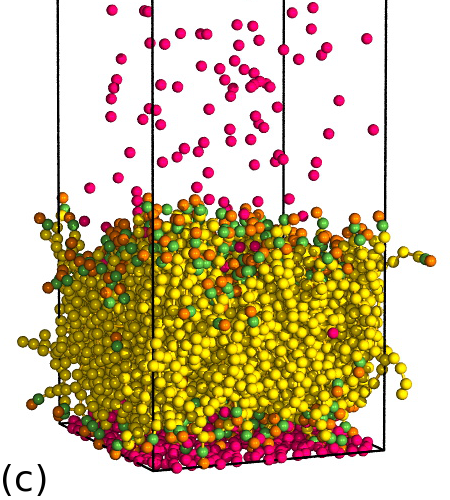}
}
\caption{\small 
Snapshots for the system with zwitterionic lipid head groups, $n=12$, $q_{\rm h}=\pm e,~\epsilon_{{\rm h}\alpha} =0$, simulated in $NVT$ ensemble, 
$a_{xy}=17.7$~\AA$^2$. The surface charge density is varied ($e/$\AA$^2$): (a) $\sigma = 0$, (b) $-0.05$, (c) $-0.1$. 
For clarity anions are not shown.
}
\label{Snaps_long_charged2}
\end{figure}

\noi {\bf Simulations for longer lipid chains.~} As a measure to verify 
that the phenomena persist for lipid chains with longer hydrophobic tails, we performed a number of simulations under similar 
conditions for zwitterionic lipids with $n=12$, $\epsilon_{\rm tt}=1.0~k_{\rm B}T$ and $a_{xy}=17.7$~\AA$^2$, where the reduction in 
tail-tail interaction parameter compensates for the increase in the chain length. The snapshots given in Fig.~\ref{Snaps_long_charged2} 
illustrate the monolayer structure in this case at three different simulation stages. Apart from supporting the above discussion, 
these snapshots also highlight the mechanism of both lipid flipping and counterion penetration through the lipid film. 
Apparently, the two processes are coupled, proceeding via formation of ion channels where ions moving along the electrostatic 
gradient are surrounded by charged lipid head groups, see Fig.~\ref{Snaps_long_charged2}(b). Clearly, this mechanism greatly 
reduces the energetic barriers associated with both processes in two ways: (i) by maximising the lipid head-ion electrostatic 
interactions and (ii) by reducing the number of contacts between ions and lipid hydrocarbon tails. 

Looking ahead, we performed several additional simulations for zwitterionic lipids composed of twelve beads with 
a larger size of the bead (up to 4.5 \AA) and a smaller step of the surface charge variation. In these test 
runs the phenomena discussed above remained essentially the same, whereas the calculated dependence of the surface potential, 
$\phi_0\equiv\phi_{z=0}$, on the surface charge density exhibited an interesting hysteresis feature (data not shown here). 
Initially, at low values of $\sigma$, the electrode potential raised linearly with increasing $\sigma$. However, at 
$\sigma^\ast\approx -0.02~e/$\AA$^2$ a sudden drop in the magnitude of $\phi_0$ was found, which coincided with a relatively 
abrupt desorption of the lipid film and an establishment of an electric double layer by the surface. When the surface charge 
was further increased, the monolayer eventually transformed into a symmetric bilayer, just as we saw in the present study, 
and the surface potential started growing again, gaining the magnitude beyond that obtained at $\sigma^\ast$. This back 
and forth going level of the surface potential in a narrow range of $\sigma$ can be rationalised as a simulation 
evidence of the first peak in $C(V)$. It is clear that in a voltammetry experiment one cannot observe such a hysteresis 
in $\phi_0(\sigma)$, because the surface charge is not set but, instead, is defined by the applied electric potential. 
Therefore, the voltage hysteresis loop detected in simulation is absent in experiment, being simply replaced by a sudden raise 
in the surface charge, as was reported, for instance, by Bizzotto and Nelson~\cite{Bizzotto-Nelson:98}, see Fig.~1 
in their paper. 

\section{Concluding remarks}
\label{Remarks}

In this work we have modelled numerically phase phenomena in lipid films adsorbed onto an electrode to which 
a varying electric potential is applied -- the setup that presents difficulties for structural analysis in experiment. 
As we have seen above, both mean-field calculations and Monte Carlo simulations predict electrostatically driven desorption 
of the lipid monolayer as the electrode surface charge density is raised beyond a certain threshold value. The prominent 
observation is that the monolayer desorption is governed by strong competition between the electrode counterions and the lipid 
hydrocarbon tails that are initially hydrophobically attached to the surface. Counterion accumulation in the vicinity 
of the progressively charged-up surface appears to be only weakly affected by the presence of the lipid film, 
implying that the normal electric double layer is eventually established, provided sufficiently long relaxation time and 
the surface charge generating a steep electrostatic gradient (initially in the absence of the counterions by the surface). 

Our MC simulations also revealed two possible scenarios of events upon displacement of the lipid monolayer from the electrode. 
In the first case, when polar groups in the lipid heads can be considered relatively inert with respect to the applied 
electric field, the desorbed monolayer readily transforms into a bilayer which then quickly departs from the surface 
into the solution. In the second case, when the lipid head group is zwitterionic (or strongly polar), the monolayer-bilayer 
transformation takes longer time, even though the initial zero-tension monolayer appears to be laterally expanded, i.e. 
with larger area per lipid than that in the first case. We note that such an in-plane expansion should be attributed 
to the structure brought about by electrostatic correlations within the head-group region of a lipid layer. This should 
also explain the relatively larger kinetic barrier and the longer time of flipping a lipid with its head group from 
one side of the film to the other. Nevertheless, charges in the lipid head groups oriented towards 
the electric double layer reside in the region of non-trivial electrostatic gradient where they appear to be trapped: 
those with the charge opposite to that of an electrode mix with the electrode counterions in the immediate proximity 
of the surface, and those with same charge as the electrode stick out and form another layer next to the inner one. 
By analogy with the term ``electric double layer'', one can in this case think of the emerging alternating charge layers 
as an ``electric triple layer'', or even an ``electric multilayer'' -- taking into account the charges within the outer 
side of the so readsorbed bilayer. It is then reasonable to assume that the emergence of these charged layers 
in a stepwise fashion is manifested in the sharp $C(V)$ peaks observed in voltammetry experiments. 
It should also be possible to verify this hypothesis in simulation by gradually scanning the relevant range 
of the surface charge density and calculating the corresponding electric potential and the interface capacitance. 

\vskip 16pt

{\large \bf Acknowledgments}

\vskip 8pt

A.~B. expresses his gratefulness to Alexandre Vakourov (CMNS, University of Leeds, UK) 
for many helpful and inspiring discussions on the electrochemical problem and experiments. 
The work of A.~B. and S.~A. was supported by the EPSRC-GB Grant No. EP/G026165/1. 

\bibliography{ABrefs}{}
\bibliographystyle{aip}

\end{document}